\documentclass[amsfonts, amssymb, amsmath, reprint]{revtex4-1}
\usepackage[english]{babel}
\usepackage[utf8]{inputenc}
\usepackage{amsthm}
\usepackage{mathtools}
\usepackage{physics}
\usepackage{xcolor}
\usepackage{graphicx}
\usepackage{adjustbox}
\usepackage{placeins}
\usepackage[T1]{fontenc}
\usepackage{csquotes}
\usepackage{amsmath}
\usepackage{bm}
\usepackage{siunitx}
\usepackage{tabularx}
\usepackage{booktabs}
\usepackage{array}

\providecommand{\dd}{\mathrm{d}}

\usepackage[pdftex, pdftitle={Mode-dependent photothermal responsivity mapping of tensile-stressed nanomechanical resonators}, pdfauthor={Thomas M. Tropper, Silvan Schmid}]{hyperref}
\newcommand{\Pabs}{P_{\mathrm{abs}}}
\newcommand{\avg}[1]{\left\langle #1 \right\rangle}

\begin{document}

\title{Mode-dependent photothermal responsivity mapping of tensile-stressed nanomechanical resonators}

\author{Thomas M. Tropper}
\email[Correspondence email address: ]{thomas.tropper@tuwien.ac.at}
\author{Silvan Schmid}
\affiliation{Institute of Sensor and Actuator Systems, TU Wien, Gusshausstrasse 27-29, 1040 Vienna, Austria.}

\date{July 31, 2026}

\begin{abstract}
    The photothermal responsivity of a tensioned nanomechanical resonator, the fractional frequency shift per absorbed optical power, depends on which vibrational mode is tracked and where the heat is deposited. We derive a closed-form model of this dependence for square membranes by treating the membrane as a grid of independent tensile fibers, each responding to its mean temperature. The responsivity map splits into a mode-independent conduction envelope and a mode term that peaks on the antinodal rows and columns of the mode, not at the antinodes, because fibers rather than points carry the modal strain energy. Scanning photothermal measurements of the $(1,1)$, $(1,2)$, $(2,1)$, and $(2,2)$ modes of a square silicon nitride membrane confirm the predicted maps with a correlation $r \geq 0.998$, with the envelope amplitude as the only fitted parameter. On trampoline resonators, the tether-limited thermal conductance flattens the envelope, making the mode term directly visible in raw scans. The measured mode-resolved maps provide the design input for dual-mode photothermal sensing with common-mode drift suppression, identifying pairs, such as a membrane's $(1,1)$/$(2,2)$ or a trampoline's torsional modes, as candidates that combine large differential responsivity with a shared thermal and mechanical environment.
\end{abstract}


\maketitle

\section{Introduction}

Nanomechanical photothermal sensing exploits a tensioned resonator as a mechanical thermometer: absorbed optical power raises the resonator temperature, relaxes the tensile stress, and detunes the resonance frequency~\cite{Schmid2023book,West2023}. This transduction principle has enabled single-molecule absorption microscopy~\cite{Chien2018}, photothermal infrared spectroscopy~\cite{Yamada2013,Kurek2017,Luhmann2023,timarac2026picogram,surdu2026quantifying}, and uncooled infrared and terahertz detection approaching fundamental noise limits~\cite{Piller2023,Zhang2024,Martini2025}.

The central design quantity is the relative power responsivity, the fractional frequency shift per absorbed power. For the fundamental mode, Kanellopulos \emph{et al.} introduced the \emph{mean temperature framework} (MTF), showing that the responsivity is governed by the \emph{mean} temperature rise of the resonator rather than by its peak temperature and used it to compare strings, drumheads, and trampolines~\cite{Kanellopulos2025}. That the response is mode-dependent has been demonstrated and exploited: Martini \emph{et al.} demonstrated by finite-element simulations the mode-dependence of the $(1,1)$, $(2,2)$, and $(3,3)$ of a square SiN membrane resonator for a large IR spot~\cite{Martini2025}, and Sadeghi \emph{et al.} showed that localizing a phononic-crystal defect mode on top of the photothermally-induced temperature field enhances the responsivity by up to an order of magnitude, directly evidencing the role of the overlap between temperature field and mode shape~\cite{Sadeghi2020}. What has been missing is a closed-form theory of this position- and mode-dependence together with a mode-by-mode experimental test.

A timely application sharpens this question. Brown \emph{et al.} recently reached thermomechanically limited frequency stability below $5\times10^{-9}$ fractional Allan deviation by tracking two mechanical modes simultaneously and subtracting their fractional frequencies, which cancels drifts common to both modes~\cite{Brown2026}. They explicitly propose extending this dual-mode scheme to thermal sensing: a mode pair with a \emph{differential} response to the localized signal heat but a \emph{common} response to global drifts would combine high responsivity with intrinsic drift rejection~\cite{Brown2026,Gavartin2013}. Designing such a pair requires precisely the quantity derived and measured here: the photothermal responsivity of individual modes as a function of the heating position.

In this work, we derive the responsivity map of a square membrane under point-like heating in closed form (Sec.~\ref{sec:theory}; full derivation in the Supplementary Material). The key physical input is a fiber picture of thermoelastic stress relaxation: the membrane acts as a dense grid of independent tensile fibers, each of which responds to its mean temperature. The map splits exactly into a mode-independent conduction envelope and a closed-form mode term that peaks on the antinodal rows and columns of the mode. We validate the model with position-resolved photothermal measurements of the $(1,1)$, $(1,2)$, $(2,1)$, and $(2,2)$ modes of a SiN membrane and extend the comparison to trampoline resonators, where the flat tether-dominated thermal envelope makes the mode term directly visible in raw scans (Secs.~\ref{sec:methods}, \ref{sec:results}). Finally, we discuss resonator geometries for common-mode-suppressed photothermal sensing (Sec.~\ref{sec:cms}).

\section{Theory}
\label{sec:theory}

To interpret spatially resolved photothermal frequency maps, we model the resonator as a square membrane of side length $L$, thickness $h$, biaxial tensile prestress $\sigma_0$, Young's modulus $E$, thermal expansion coefficient $\alpha$, and thermal conductivity $\kappa$, clamped at $T_0$ along its boundary. Its out-of-plane eigenmodes are
\begin{equation}
    \begin{split}
        &\phi_{nm}(x,y) = \sin\!\left(\frac{n\pi x}{L}\right)\sin\!\left(\frac{m\pi y}{L}\right),\\
        &\omega_{nm} = \sqrt{\frac{\sigma_0}{\rho}}\,\frac{\pi}{L}\sqrt{n^2+m^2}.
        \label{eq:modes-main}
    \end{split}
\end{equation}

A focused laser absorbing power $\Pabs$ at a point $(x_0,y_0)$ locally heats the membrane and relaxes its tension, shifting the eigenfrequency of mode $(n,m)$. The model rests on two assumptions: (i) negligible thermal conduction to the surrounding gas, i.e.\ operation in vacuum, and (ii) negligible radiative heat transfer, which becomes significant for high-aspect-ratio membrane resonators \cite{zhang2020radiative,piller2020thermal}. For the present membranes ($L=500\,$\textmu m, $h=50\,$nm), radiation is a small correction (we retain the conduction-only model throughout the main text); a linearized radiative extension of the model, which screens the low-order temperature components and slightly flattens the envelope, is derived in the Supplementary Material, Sec.~\ref{sec:SI-radiation}. Using first-order (Rayleigh) perturbation theory together with the steady-state heat equation, this shift can be written as
\begin{equation}
    \frac{\delta\omega_{nm}}{\omega_{nm}}(x_0,y_0)
    = \mathcal{R}_T\; \frac{\Pabs}{\kappa h}F_{nm}(x_0,y_0),
    \label{eq:master-main}
\end{equation}
with the temperature responsivity \cite{Schmid2023book}
\begin{equation}
    \mathcal{R}_T=-\,\frac{E\,\alpha}{2\,\sigma_0}
\end{equation}
and where $F_{nm}$ is a dimensionless responsivity map of order $10^{-1}$ (full derivation in the Supplementary Material, Sec.~\ref{sec:SI-theory}).

The key physical input is that the membrane behaves, to excellent approximation, as a dense grid of independent tensile fibers along $x$ and $y$: because tension must be uniform along each fiber in mechanical equilibrium, a fiber's tension change is set by its \emph{mean} temperature, not its local temperature (Supplementary Material, Sec.~\ref{sec:SI-fibers}). This single fact determines which parts of the membrane a given mode is sensitive to when heated.

Carrying this through, $F_{nm}$ decomposes \emph{exactly} into a smooth background and a closed-form mode term, $F_{nm} = A_{nm} + B_{nm}$ (Supplementary Material, Sec.~\ref{sec:SI-decomposition}). The background $A_{nm}$ coincides, to leading order, with the mean-temperature (MTF) envelope of the membrane~\cite{Kanellopulos2025},
\begin{equation}
    A_{nm}(x_0,y_0)\approx\frac{\kappa h}{\Pabs}\avg{\Delta T}(x_0,y_0),
    \label{eq:env-main}
\end{equation}
while the mode term has the closed form
\begin{equation}
    \begin{split}
        B_{nm}(x_0,y_0)
        &= \frac{1}{2\pi^2(n^2+m^2)}\\
        &\left[
        \frac{n^2}{m^2}\,\sin^2\!\Big(\frac{m\pi y_0}{L}\Big)\;w_{2m}\!\Big(\frac{\pi x_0}{L}\Big)
        \right.\\
        &\left. {}+ \frac{m^2}{n^2}\,\sin^2\!\Big(\frac{n\pi x_0}{L}\Big)\;w_{2n}\!\Big(\frac{\pi y_0}{L}\Big)
        \right],
        \label{eq:Bclosed-main}
    \end{split}
\end{equation}
where $w_c(\theta)$ is an edge window that equals $1$ in the interior and drops to $0$ within $\sim L/(2\pi n)$ and $\sim L/(2\pi m)$, respectively, of the clamped boundary (defined in Supplementary Material, Sec.~\ref{sec:SI-decomposition}).
Equation~\eqref{eq:Bclosed-main} captures the essential experimental signature: the mode term is largest on the \emph{antinodal rows and columns} of the mode, not at the antinodes themselves, because it is fibers, not points, that carry the modal strain energy. The background $A_{nm}$, by contrast, simply follows the thermal conductance profile of the membrane and is independent of $(n,m)$ to leading order.

\section{Methods}
\label{sec:methods}

\begin{figure}
    \centering
    \includegraphics[width=\linewidth]{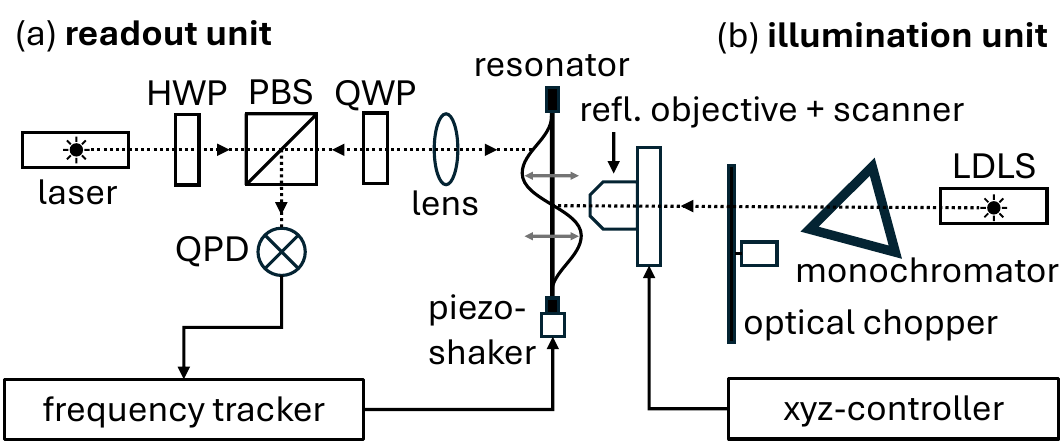}
    \caption{\textbf{Experimental setup for nanomechanical photothermal mode mapping.} (a)~\emph{Readout unit.} An optical-lever scheme --- laser, half-wave plate (HWP), polarizing beam splitter (PBS), quarter-wave plate (QWP), lens, and quadrant photodetector (QPD) --- detects the out-of-plane motion of the resonator. A frequency tracker (PHILL) drives a piezo-shaker in a self-sustaining oscillator loop to keep the resonator on resonance. (b)~\emph{Illumination unit.} A laser-driven light source (LDLS) is spectrally filtered by a monochromator, modulated on/off by an optical chopper, and focused onto the resonator by a reflective objective; an $xyz$ nanopositioner scans the focal spot across the surface.
    }
    \label{fig:experimental-setup}
\end{figure}

The experiment comprises two subsystems: a readout unit that detects the mechanical motion, and an illumination unit that deposits a focused, wavelength-selected optical power at a controlled position on the resonator (Fig.~\ref{fig:experimental-setup}). All measurements were performed on a square SiN membrane and a trampoline resonator mounted inside a vacuum chamber maintained below $10^{-5}$\,hPa to suppress gas damping and convective heat transfer~\cite{Verbridge2008}.

\subsection{Readout unit}

The resonator's out-of-plane motion was detected via an optical lever using a $785$\,nm laser (LDM785, Thorlabs, Inc.), chosen for its weak absorption in SiN to minimize photothermal loading by the readout itself. The laser power at the sample was $P_\mathrm{readout}\approx170$\,\textmu W. The reflected beam was detected with a quadrant photodetector (QPD) (PD4Q-100k, Koheron). The resonance frequency of the mode under study was tracked in real time with a frequency-tracking unit (PHILL, Invisible-Light Labs GmbH) operated as a self-sustaining oscillator (SSO): the detected motion is band-pass filtered, phase-shifted, amplified, and fed back to a piezoelectric actuator beneath the chip, so that the loop continuously oscillates at the mechanical resonance and the instantaneous frequency is available as the measurement output~\cite{Besic2023}. Compared to a phase-locked loop, the SSO requires no external reference oscillator and its tracking bandwidth is not limited by a phase-detector or feedback control time constants~\cite{Besic2023}.

\subsection{Illumination unit}

Local optical excitation was provided by a broadband light source (XWS-65, ISTEQ) coupled to an integrated monochromator (Hyperchromator II, Mountain Photonics Gmbh) for wavelength selection. The output beam was collimated, intensity-modulated with an optical chopper, and focused onto the resonator surface with a reflective objective (5006-190, Beck Optronic Solutions) of numerical aperture $\mathrm{NA}=0.65$ and working distance $d=1.8$\,mm, which avoids chromatic aberration across the source spectrum, yielding a focal spot size of $30$\,\textmu m in diameter. The spot size was determined by knife-edge measurement and is etendue limited. The objective was mounted on a $xyz$ piezo nanopositioning stage (PIMars P-56x and E-727, Physik Instrumente (PI)), enabling deterministic positioning of the focal spot across the resonator surface. The stage provides a travel range of $300$\,\textmu m.

An excitation wavelength of $\lambda=320$\,nm was chosen as a compromise between available source power and the strong interband absorption of silicon nitride in the near-UV, which maximizes the absorbed fraction of the incident power. The incident power at the sample was $P_0\approx13$\,\textmu W. As the absorptance of the SiN film at this wavelength is not independently calibrated, all maps are reported per \emph{incident} power; the (position-independent) absorptance enters only through the overall amplitude $k$ of the mode-term fit and therefore does not affect the extracted spatial pattern (Sec.~\ref{sec:SI-pipeline}). The optical chopper was operated at $8$\,Hz, effectively switching the illumination on and off rather than sinusoidally modulating it. The thermal time constant of the membranes was measured to be $\tau_\mathrm{th}\approx13$\,ms, well below the modulation time constant of $\tau_\mathrm{mod}=125$\,ms, providing full on/off cycles with no low-pass correction required. For each spot position, several on/off cycles were recorded and the resulting plateau maxima and minima were averaged to obtain the signal difference $\Delta f$. Given the short acquisition time of $0.5$\,s per measurement, thermal drift is negligible on this timescale, and the signal-to-noise ratio at this wavelength is sufficiently high that this direct min/max estimator is adequate; lock-in demodulation, typically employed to recover small signals from noisy or drift-prone backgrounds, was therefore not required.

\subsection{Measurement protocol}
For each membrane the modes $(1,1)$, $(1,2)$, $(2,1)$ and $(2,2)$ were addressed sequentially by locking the SSO to the respective resonance, and the heating spot was raster-scanned across the surface on a $15$\,\textmu m grid; the full-membrane maps were assembled from five overlapping tiles as described in Sec.~\ref{sec:SI-pipeline}. The mode-specific contribution was then obtained by subtracting the modelled conduction envelope and normalizing by its peak, $B_{nm}/A_\mathrm{max}$, rather than by dividing by a separately measured reference mode; a single overall amplitude is the only fitted quantity (Sec.~\ref{sec:SI-pipeline}). For the trampoline, single tiles covering the central pad were recorded for the tether-dominated, torsional and pad-localized modes; no stitching or envelope subtraction is required, since the tether-limited conductance renders the envelope nearly position-independent (Sec.~\ref{sec:results-trampolines}).

\section{Results \& Discussion}
\label{sec:results}

\subsection{Square membranes}
\label{sec:results-membranes}

\begin{figure*}
    \centering
    \includegraphics[width=0.9\textwidth]{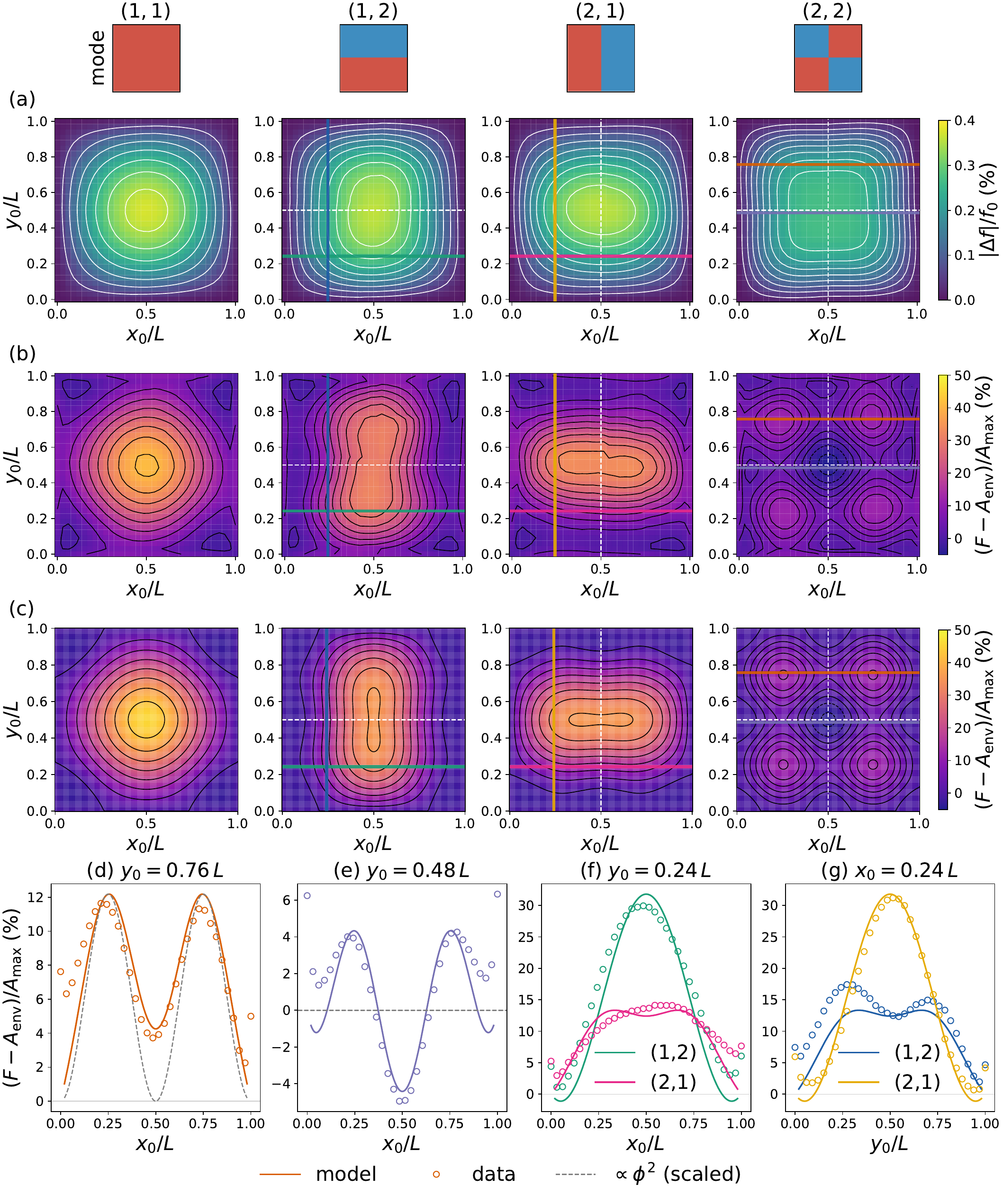}
    \caption{\textbf{Mode-resolved photothermal maps of a square SiN membrane with side length $L=500$\,\textmu m, thickness $h=50$\,nm and a tensile stress of $\sigma_0=230$\,MPa.} (a) Raw relative frequency-response maps ($|\Delta f|/f_0$) of a stitched full-membrane scan for the $(1,1)$, $(1,2)$, $(2,1)$ and $(2,2)$ modes, at resonance frequencies of $325$, $513$, $515$ and $650$\,kHz, respectively. (b) Mode contribution $B_{nm}/A_\mathrm{max}$, obtained by subtracting the conduction-dome envelope $A_\mathrm{env}$ and normalizing by its maximum $A_\mathrm{max}$ (at the membrane center). (c) Corresponding fiber-model predictions. (d-g) Line cuts through the maps: along an antinodal row (d) and a nodal row (e) of the $(2,2)$ mode, and along a common horizontal row ($y_0 \approx 0.25\,L$) and ($x_0 \approx 0.25\,L$) through the $(1,2)$ and $(2,1)$ modes (f) \& (g), respectively. Circles: data; solid lines: conduction-only fiber model.
    }
    \label{fig:membranes}
\end{figure*}

Figure~\ref{fig:membranes} compares the measured photothermal response of the four lowest modes with the fiber model. We first note a counterintuitive feature of the raw maps [Fig.~\ref{fig:membranes}(a)]: all four modes are dominated by the same broad dome peaking at the membrane center, regardless of their mode shape. This is most striking for the $(2,2)$ mode, whose center is a nodal crossing: the raw response is nonetheless maximal there. The fiber model reproduces this quantitatively. For the ideal membrane, the global maximum of $F_{22}$ lies at the center ($F_{22}=0.0704$), and the mode-specific structure is only a $\sim\!\pm20\%$ modulation on top of the conduction dome. Physically, the response follows the conduction envelope $A_\mathrm{env}$, i.e.\ the mean temperature rise, which peaks for central heating irrespective of the mode; the mode enters only through the smaller term $B_{nm}$. The measured raw maxima of all four modes coincide with the membrane center to within one scan pixel, confirming this picture.

Subtracting the modeled conduction envelope and normalizing by its peak isolates the mode term. Figure~\ref{fig:membranes}(b) shows the measured $B_{nm}/A_\mathrm{max} = (F_{nm}-A_\mathrm{env})/A_\mathrm{max}$ maps, in close agreement with the fiber-model prediction [Fig.~\ref{fig:membranes}(c)]. The distinguishing feature is that the maxima lie on the antinodal \emph{rows and columns} of each mode --- not at the displacement antinodes. This is made explicit by the line cuts in Fig.~\ref{fig:membranes}(d)--(g): along the $(2,2)$ antinodal row the fiber model tracks the data across the full profile, whereas the naive local model ($\propto\phi_{nm}^2$) predicts sharp peaks at the antinodes that the data do not show; along the $(2,2)$ nodal row the data follow the fiber model's structured, non-zero response, while the naive model is identically zero there. These two cuts alone discriminate decisively between the fiber and naive-local descriptions.

The doublet $(1,2)$/$(2,1)$ behaves like the other modes: panels (f) \& (g) show representative line cuts through each, again in agreement with the fiber model. The two maps are related by a reflection across the diagonal, as expected for a transposed mode pair.

Across all four modes, the conduction-only fiber model reproduces the raw maps with a correlation $r \geq 0.998$ and a residual of $1.0$--$1.6\%$ of the peak response. Following the fit strategy of the Supplementary Material (Sec.~\ref{sec:SI-limits}), only the envelope amplitude is floated --- within the Poisson-ratio band of order $30\%$ --- while the mode-term shape is fixed and free of adjustable parameters.

Several small deviations remain, all understood. The dominant one is a reproducible $\sim\!1\%$ flattening of the dome relative to the conduction-only prediction, consistent in both sign and magnitude with the onset of radiative heat transport expected at this membrane size for $50$\,nm SiN; we do not correct for it here, but a linearized radiative extension that reproduces this flattening is derived in the Supplementary Material (Sec.~\ref{sec:SI-radiation}). Finite laser-spot smoothing affects only the immediate vicinity of the frame and is negligible over the scanned interior. Finally, the doublet $(1,2)$/$(2,1)$ is not exactly degenerate on a physical membrane: the small frequency splitting ($513$ vs.\ $515$\,kHz) reflects a slight deviation from a perfect square. We therefore assign these two modes from their measured spatial patterns rather than from their frequencies, and the same residual splitting sets the common-mode-rejection floor discussed in Sec.~\ref{sec:cms}.

\subsection{Trampoline}
\label{sec:results-trampolines}

\begin{figure*}
    \centering
    \includegraphics[width=0.9\textwidth]{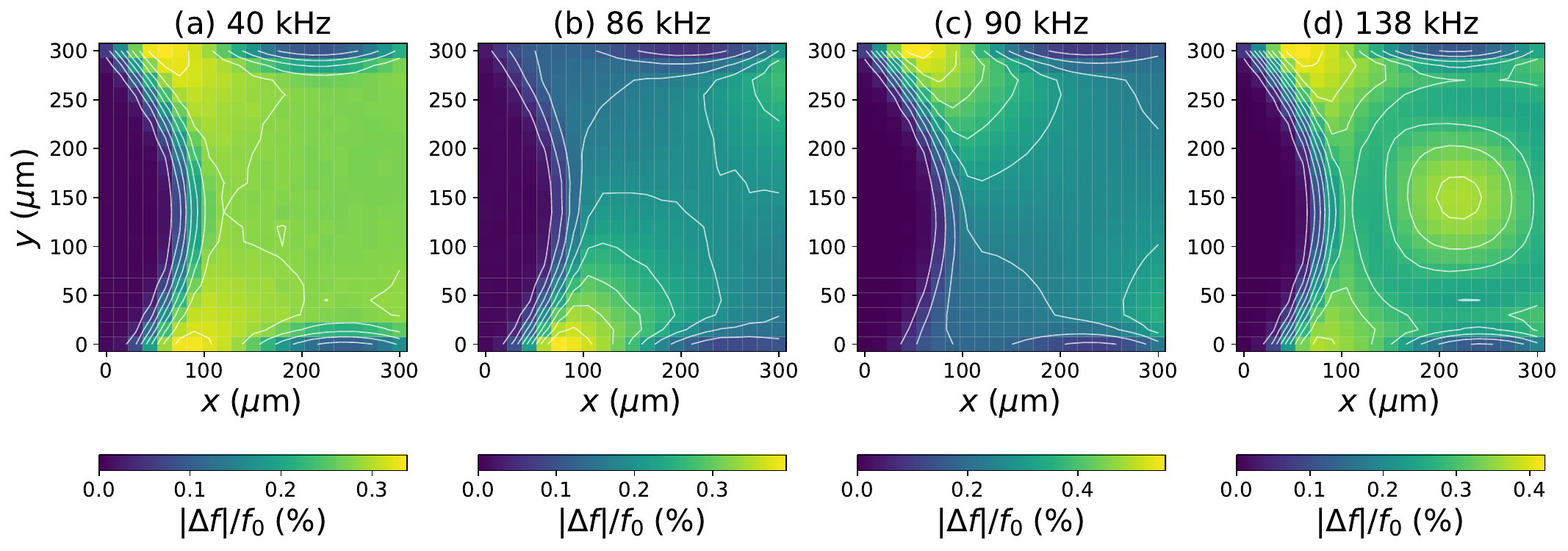}
    \caption{\textbf{Photothermal maps of trampoline modes of a trampoline with a window length of $L_\mathrm{window}=1$\,mm, a tether length of $L_\mathrm{tether}\approx500$\,\textmu m and a pad side length of $L_\mathrm{pad}\approx300$\,\textmu m with a nominal tensile stress of $\sigma_0\approx40$\,MPa.} Raw relative frequency-response maps ($|\Delta f|/f_0$) over the central pad; no envelope correction is applied, since the tether-dominated thermal conductance makes the conduction background nearly position-independent (unlike the membrane, cf.\ Fig.~\ref{fig:membranes}). (a) Tether-dominated fundamental mode ($40$\,kHz). (b, c) The two orthogonal torsional modes ($86$ and $90$\,kHz). (d) Pad-localized mode ($138$\,kHz).
    }
    \label{fig:trampolines}
\end{figure*}

Trampolines provide a clean test of the decomposition $F = A + B$: because the thermal conductance is set by the tethers, the envelope $A$ is nearly independent of the heating position on the pad, so the raw scan directly displays the mode term $B$. Figure~\ref{fig:trampolines} compares four cases: (a) a tether-dominated (fundamental) mode at $40$\,kHz, (b) and (c) two orthogonal torsional modes at $86$ and $90$\,kHz, and (d) a pad-localized mode at $138$\,kHz.

The fundamental mode Figure~\ref{fig:trampolines}(a) shows the predicted flat pad response within a standard deviation of $\sigma<2.3\%$ of the mean response (peak-to-peak $11.1\%$), as the quantitative measure of a featureless pad. The tethers constrain and define the thermal conductance, and all absorbed heat must exit through them; since each tether relaxes only in proportion to its \emph{mean} temperature, the response is insensitive to where on the pad the heat is deposited.

The two torsional maps Figure~\ref{fig:trampolines}(b) and (c) are clearly distinct, with mutually orthogonal symmetry axes. These patterns can be understood by picturing the trampoline as two crossed strings (fibers) connected by a central pad: for each torsional mode one string undergoes a bending motion while the orthogonal string follows in torsion. Within the fiber picture the responsivity is then expected to increase when the bending string is heated and to decrease when the torsional string is heated, consistent with the observed sign structure.

The pad-localized mode Figure~\ref{fig:trampolines}(d) shows a single centered, circularly-symmetric feature that is visible directly in the raw scan without any envelope correction. The flat pad response $A$ makes the mode-dependent term $B$ appear without any data processing.

\subsection{Differential mode signals}
\label{sec:cms}

\begin{figure*}
    \centering
    \includegraphics[width=0.9\textwidth]{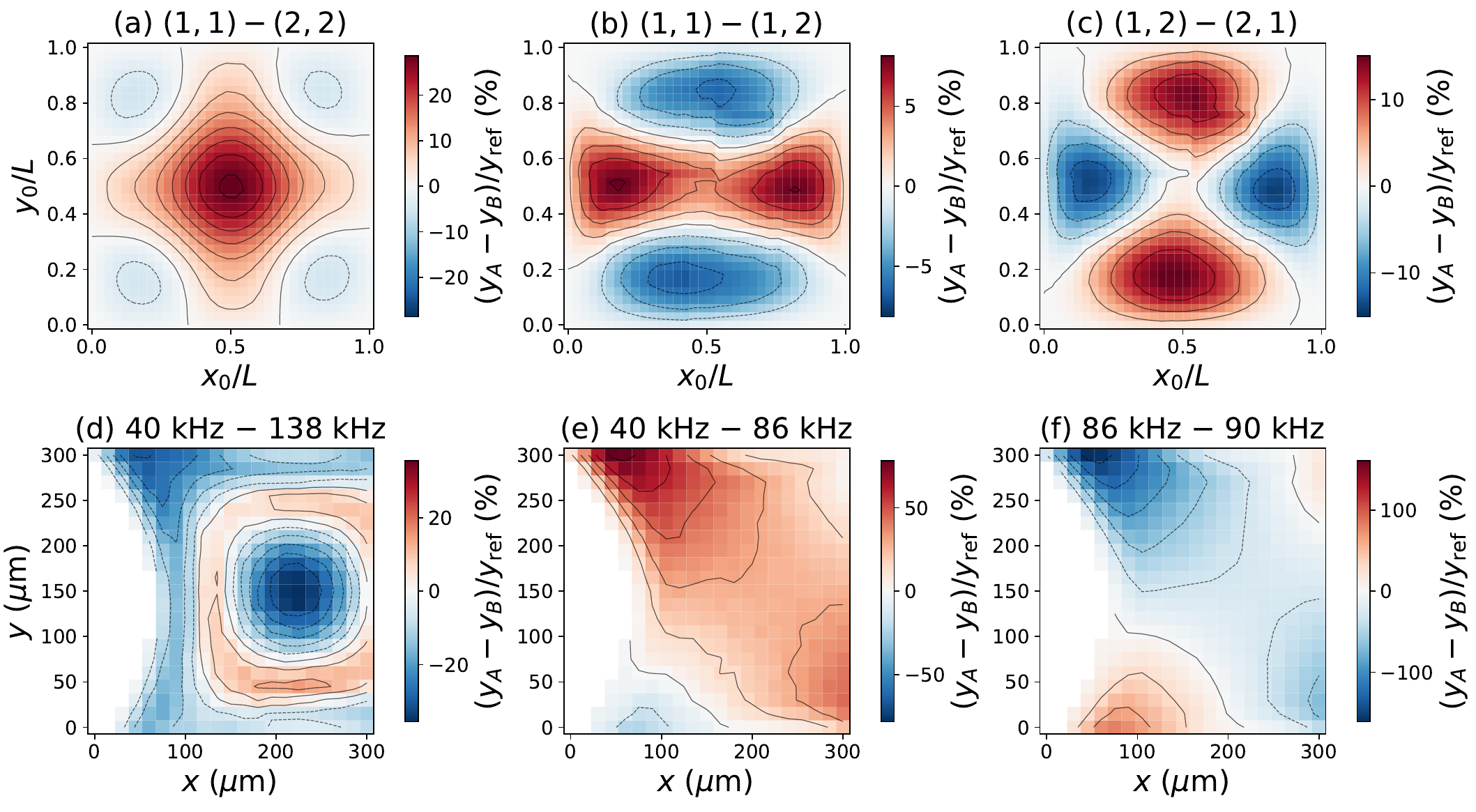}
    \caption{\textbf{Normalized differential relative frequency-response maps for selected membrane and trampoline mode pairs.} The relative frequency-response maps ($\mathrm{y}=|\Delta f|/f_0$) of two modes ($\mathrm{y}_A, \mathrm{y}_B$) are subtracted and normalized by a common reference $\mathrm{y_{ref}}$. For the membrane pairs (a--c), $\mathrm{y_{ref}}=\max(\mathrm{y}_{(1,1)})$, the peak (central) response of the $(1,1)$ mode. For the trampoline pairs (d--f), $\mathrm{y_{ref}}$ is the central response of the tether-dominated $40$\,kHz mode.
    }
    \label{fig:cmr}
\end{figure*}

Dual-mode common-mode subtraction cancels correlated noise and parameter variations experienced by both tracked modes, while a differential signal survives~\cite{Brown2026,Gavartin2013}. An ideal photothermal mode pair, therefore, maximizes the differential responsivity to the localized signal while equally coupling to the same thermal bath, temperature, and environmental fluctuations.
Figure~\ref{fig:cmr} shows the differential maps for membrane and trampoline mode pairs normalized to a common reference, which is the maximal responsivity obtained at the center of the respective fundamental modes.

For the square membrane, the maximal differential signal is obtained in the membrane center for the (1,1)/(2,2) mode pair, reaching 20\% signal strength compared to the (1,1) mode alone (Fig.~\ref{fig:cmr}a). The (1,2)/(2,1) doublet provides a lower differential signal of approximately 10\% (Fig.~\ref{fig:cmr}c). However, the two mode frequencies typically are close and can overlap in the case of a degeneracy, making simultaneous tracking of both challenging.

In trampolines, while the mode pairs consisting of the fundamental mode (common reference) and a torsional or localized pad mode show a differential signal of up to 50\% (Fig.~\ref{fig:cmr}d\&e), the orthogonal torsional mode pair produces a differential signal larger than the common reference (Fig.~\ref{fig:cmr}f). The potential drawback of the latter pair is that the two modes live on orthogonal fibers, whose clamping points to the frame are also orthogonal and will likely experience uncorrelated thermomechanical fluctuations.

\section{Conclusion}
The paper derives and experimentally validates a closed-form model for the position- and mode-dependent photothermal responsivity of tensioned nanomechanical resonators. The key physical idea is a fiber picture: a square membrane behaves as a grid of independent tensile fibers, each relaxing in proportion to its mean temperature. This yields an exact decomposition of the responsivity map $F_{nm} = A_{nm} + B_{nm}$, where $A$ is a mode-independent conduction envelope (recovering the mean-temperature framework of Kanellopulos et al. \cite{Kanellopulos2025}) and $B$ is a closed-form mode term that peaks on the antinodal rows and columns rather than at the antinodes, since fibers, not points, carry the modal strain energy. Scanning photothermal measurements on a 500~\textmu m SiN membrane confirm the model for the (1,1), (1,2), (2,1), and (2,2) modes with raw-map correlation $r\geq0.998$  and residuals of 1.0--1.6\%, with only the envelope amplitude as a fitted parameter; the dominance of the central conduction dome even for the (2,2) mode (nodal crossing at the center) is reproduced quantitatively. On trampolines, the tether-limited thermal conductance flattens the envelope on the central pad, so the mode term is directly visible in raw scans: featureless for the tether-dominated fundamental, structured with orthogonal symmetry axes for the torsional pair, and directly displaying the mode shape for the pad-localized mode. The practical payoff is a design tool for dual-mode common-mode-suppressed photothermal sensing \cite{Brown2026}: differential maps identify harmonic pairs, in particular (1,1)/(2,2) membrane and the trampoline torsional pairs, as the most promising candidates, with differential signals from ${\sim}20\%$ up to exceeding the fundamental-mode reference. The ideal mode pair for common-mode suppression requires further study, as it depends on the common coupling of both modes to the relevant noise and fluctuation sources. In that sense, the present maps identify candidates rather than a definitive choice; measuring the correlated frequency fluctuations of the candidate pairs is the natural next step.

\begin{acknowledgments}
    We thank Cindy Regal and Sofia C. Brown from the University of Colorado for the discussions that inspired us to pursue this research. We thank Sophia Schneider for her support in the cleanroom and for fabricating the samples. The research received funding from the Novo Nordisk Foundation under grant no. NNF22OC0077964 - MASMONADE. 
\end{acknowledgments}

\section*{Data availability}
The scan data and the analysis and stitching code (Sec.~\ref{sec:SI-pipeline}) are available from the corresponding author on reasonable request.

\bibliography{refs}


\clearpage
\setcounter{section}{0}
\setcounter{equation}{0}
\setcounter{figure}{0}
\setcounter{table}{0}
\renewcommand{\thesection}{S\arabic{section}}
\renewcommand{\theequation}{S\arabic{equation}}
\renewcommand{\thefigure}{S\arabic{figure}}
\renewcommand{\thetable}{S\arabic{table}}

\onecolumngrid
\begin{center}
  \textbf{\large Supplementary Material}\\[2pt]
  \textit{Mode-dependent photothermal responsivity mapping of tensile-stressed nanomechanical resonators}
\end{center}
\vspace{1em}

\section{Derivation of the photothermal responsivity map}
\label{sec:SI-theory}

This section gives the complete, step-by-step derivation of Eq.~\eqref{eq:master-main} in the main text, in five steps: (i) first-order frequency perturbation theory, (ii) the steady-state temperature field of a point heat source, (iii) the fiber picture of thermoelastic stress relaxation, (iv) assembly into the responsivity map $F_{nm}$, and (v) its exact closed-form decomposition. This model was derived with the help of Claude Fable 5 from Anthropic.

\subsection{First-order frequency perturbation}
\label{sec:SI-rayleigh}

For a tension-dominated membrane (bending negligible for $\sigma_0\gtrsim$~MPa at these dimensions), the eigenfrequency follows the Rayleigh quotient
\begin{equation}
  \omega_{nm}^2
  = \frac{\displaystyle\int \sigma_{ij}\,\partial_i\phi\,\partial_j\phi\;\dd A}
         {\displaystyle\int \rho\,\phi^2\;\dd A},
  \qquad i,j\in\{x,y\}.
  \label{eq:SI-rayleigh}
\end{equation}
Because this expression is stationary with respect to variations of $\phi$ around an eigenmode, a small prestress perturbation $\sigma_{ij}=\sigma_0\delta_{ij}+\delta\sigma_{ij}(\mathbf r)$ shifts the eigenvalue to first order \emph{without} requiring the (also perturbed) mode shape:
\begin{equation}
  \frac{\delta\omega}{\omega}
  = \frac{1}{2}\,
    \frac{\int \delta\sigma_{ij}\,\partial_i\phi\,\partial_j\phi\,\dd A}
         {\sigma_0\int|\nabla\phi|^2\,\dd A}.
  \label{eq:SI-pert}
\end{equation}
The stress perturbation is sampled by the mode's \emph{strain density}
$\partial_i\phi\,\partial_j\phi$, not by the displacement $\phi^2$. For the modes of
Eq.~\eqref{eq:modes-main},
\begin{equation}
  \int_0^L\!\!\int_0^L |\nabla\phi|^2\,\dd x\,\dd y = (a^2+b^2)\,\frac{L^2}{4},
  \qquad a=\frac{n\pi}{L},\ b=\frac{m\pi}{L}.
  \label{eq:SI-norm}
\end{equation}

\subsection{Temperature field of a point source}
\label{sec:SI-temperature}

In steady state, with the frame held at $T_0$ and radiation neglected (justified in Sec.~\ref{sec:SI-limits}), the temperature rise $\Delta T=T-T_0$ obeys
\begin{equation}
  -\kappa h\,\nabla^2 \Delta T = \Pabs\,\delta(x-x_0)\,\delta(y-y_0),
  \qquad \Delta T = 0 \text{ on the boundary}.
  \label{eq:SI-heat}
\end{equation}
Expanding in the Dirichlet basis $\sin(p\pi x/L)\sin(q\pi y/L)$ gives the standard 2D Green's function
\begin{equation}
  \Delta T(x,y) = \frac{4\Pabs}{\pi^2 \kappa h}
  \sum_{p,q\ge1}
  \frac{\sin\frac{p\pi x}{L}\sin\frac{q\pi y}{L}\;
        \sin\frac{p\pi x_0}{L}\sin\frac{q\pi y_0}{L}}{p^2+q^2}.
  \label{eq:SI-T}
\end{equation}
This diverges logarithmically at the source (the usual spreading resistance), regularized in practice by the finite laser spot size.

\subsection{From temperature to stress: the fiber picture}
\label{sec:SI-fibers}

\emph{A tempting but incorrect shortcut}: Writing $\delta\sigma_{ij}(\mathbf r)=-E\alpha\,\Delta T(\mathbf r)\,\delta_{ij}$ (``hot spots locally lose tension'') is \emph{not} in mechanical equilibrium ($\partial_j\delta\sigma_{ij}\neq0$). Inserted into Eq.~\eqref{eq:SI-pert}, it predicts the strongest response near the clamped edges, contradicting measured maps, which peak in antinodal regions. Elastic redistribution to restore equilibrium changes the answer qualitatively.

We instead treat the membrane as a dense grid of tensile fibers along $x$ and $y$, with uniform tension change per fiber:
\begin{equation}
  \delta\sigma_{xx}(x,y) = \delta\sigma_x(y), \qquad
  \delta\sigma_{yy}(x,y) = \delta\sigma_y(x), \qquad
  \delta\sigma_{xy} = 0,
  \label{eq:SI-ansatz}
\end{equation}
which satisfies $\partial_j\delta\sigma_{ij}=0$ identically. Appendix~\ref{app:projector} shows this reproduces the \emph{exact} plane-stress thermoelastic solution in the membrane interior.

For a single $x$-fiber at height $y$ (1D elastic line, fixed ends), static equilibrium requires uniform stress along the fiber. Enforcing zero net elongation ($u(L)-u(0)=0$) gives
\begin{equation}
  \boxed{\;
  \delta\sigma_x(y) = -E\alpha\,\overline{\Delta T}_x(y),
  \qquad
  \overline{\Delta T}_x(y) \equiv \frac{1}{L}\int_0^L \Delta T(x,y)\,\dd x,
  \;}
  \label{eq:SI-fiber}
\end{equation}
and symmetrically for $\delta\sigma_y(x)$. \textbf{This is the essential physics of the whole problem}: a fiber's tension responds only to its mean temperature, so heating a point $(x_0,y_0)$ relaxes predominantly the one row and one column of fibers passing through it.

\subsection{Assembly: which fibers does the mode load?}
\label{sec:SI-assembly}

Inserting Eq.~\eqref{eq:SI-ansatz} into the numerator of Eq.~\eqref{eq:SI-pert} and grouping by fibers gives
\begin{equation}
  \frac{\delta\omega}{\omega}
  = \frac{a^2\!\int_0^L \delta\sigma_x(y)\sin^2(by)\,\dd y
        + b^2\!\int_0^L \delta\sigma_y(x)\sin^2(ax)\,\dd x}
         {\sigma_0\,(a^2+b^2)\,L}.
  \label{eq:SI-assembled}
\end{equation}
Here $\tfrac{L}{2}a^2\sin^2(by)$ is the modal tensile energy carried by the $x$-fiber at height $y$: the fibers running through the displacement antinodes carry the most modal strain energy. Combined with Eq.~\eqref{eq:SI-fiber}, the frequency responds most when heat lands on the most-loaded fibers --- i.e.\ on antinodal rows or columns, even though the exact antinode itself is a point of zero local strain.

Evaluating the two integrals in Eq.~\eqref{eq:SI-assembled} using Eqs.~\eqref{eq:SI-fiber} and \eqref{eq:SI-T} (elementary trigonometric integrals, given in full in Appendix~\ref{app:integrals}) and using $a^2+b^2=\pi^2(n^2+m^2)/L^2$ yields the master result already stated in the main text:
\begin{equation}
  \frac{\delta\omega_{nm}}{\omega_{nm}}(x_0,y_0)
  = -\frac{E\alpha \Pabs}{2\sigma_0 \kappa h}\,F_{nm}(x_0,y_0),
  \label{eq:SI-masterrepeat}
\end{equation}
\begin{equation}
  F_{nm}(x_0,y_0)
  = \frac{64\,n^2 m^2}{\pi^4\,(n^2+m^2)}
  \sum_{p,q\;\mathrm{odd}}
  \frac{\sin\frac{p\pi x_0}{L}\,\sin\frac{q\pi y_0}{L}}
       {p\,q\,(p^2+q^2)}
  \left[\frac{1}{4m^2-q^2} + \frac{1}{4n^2-p^2}\right].
  \label{eq:SI-Fseries}
\end{equation}
$F_{nm}$ is dimensionless, symmetric under $(n,x_0)\leftrightarrow(m,y_0)$, vanishes on the boundary, and its series converges absolutely (terms decay as $p^{-3}q^{-3}$, so a $9\times9$ truncation is already accurate to the percent level). $F_{nm}>0$ everywhere: heating always lowers the frequency, as expected for $\alpha>0$.

\subsection{Exact decomposition: envelope, mode term, edge windows}
\label{sec:SI-decomposition}

The series in Eq.~\eqref{eq:SI-Fseries} hides simple structure, exposed by two Fourier-sum lemmas (proofs in Appendix~\ref{app:lemmas}):

\emph{Oscillatory sum:}\label{lem:osc-SI} $\displaystyle\sum_{q\;\mathrm{odd}}\frac{\sin q\theta}{q\,(4m^2-q^2)} = \frac{\pi}{8m^2}\,\sin^2(m\theta).$

\emph{Monotone sum:}\label{lem:mono-SI}
For $c>0$: $\displaystyle\sum_{p\;\mathrm{odd}}\frac{\sin p\theta}{p\,(p^2+c^2)} = \frac{\pi}{4c^2}\,w_c(\theta)$, with window function $w_c(\theta)\equiv1-\cosh[c(\theta-\pi/2)]/\cosh(c\pi/2)$, which equals $1$ in the interior and drops to $0$ within a boundary layer of width $\sim1/c$.

Using the partial-fraction identity $\frac{1}{(p^2+q^2)(4m^2-q^2)} = \frac{1}{p^2+4m^2}\left[\frac{1}{p^2+q^2}+\frac{1}{4m^2-q^2}\right]$ (and its counterpart), $F_{nm}$ splits \emph{exactly} into a smooth part $A_{nm}$ and a mode part $B_{nm}$, $F_{nm}=A_{nm}+B_{nm}$. Lemmas~\ref{lem:osc-SI} and \ref{lem:mono-SI} evaluate the mode part in closed form:
\begin{equation}
  B_{nm}(x_0,y_0)
  = \frac{1}{2\pi^2(n^2+m^2)}
  \left[
    \frac{n^2}{m^2}\,\sin^2\!\frac{m\pi y_0}{L}\;w_{2m}\!\Big(\frac{\pi x_0}{L}\Big)
    +
    \frac{m^2}{n^2}\,\sin^2\!\frac{n\pi x_0}{L}\;w_{2n}\!\Big(\frac{\pi y_0}{L}\Big)
  \right].
  \label{eq:SI-Bclosed}
\end{equation}
In the interior ($w\to1$) this is a separable stripe pattern (\emph{not} the product $\phi_{nm}^2$), anisotropically weighted by $(n/m)^{\pm2}$, maximal on antinodal rows/columns, with amplitude decreasing as $1/(n^2+m^2)$. The windows switch the pattern off within $\sim L/2\pi m$ of the edges, enforcing $F=0$ on the frame.

The remaining smooth part,
\begin{equation}
  A_{nm}(x_0,y_0)
  = \frac{64 n^2 m^2}{\pi^4 (n^2+m^2)}
    \sum_{p,q\;\mathrm{odd}}
    \frac{\sin\frac{p\pi x_0}{L}\sin\frac{q\pi y_0}{L}}{pq\,(p^2+q^2)}
    \left[\frac{1}{p^2+4m^2}+\frac{1}{q^2+4n^2}\right],
  \label{eq:SI-A}
\end{equation}
is compared with the exact mean temperature of the membrane, $\frac{\kappa h}{\Pabs}\avg{\Delta T}(x_0,y_0) = \frac{16}{\pi^4} \sum_{p,q\,\mathrm{odd}}\frac{\sin(p\pi x_0/L)\sin(q\pi y_0/L)}{pq(p^2+q^2)}$. For the low $(p,q)$ that dominate both series, the bracket in Eq.~\eqref{eq:SI-A} tends to $(n^2+m^2)/4n^2m^2$, matching the mean-temperature prefactor exactly, so $A_{nm}\approx\frac{\kappa h}{\Pabs}\avg{\Delta T}(x_0,y_0)$ up to a mode-dependent correction that shrinks with mode order (e.g.\ $\leq4.5\%$ for $(2,2)$, numerically verified). Physically: the frequency responds to the strain-weighted mean temperature, which approaches the plain mean temperature as the mode order increases relative to the temperature field's spatial variation.

\emph{Remark:} evaluated at the membrane center, the expression for $\avg{\Delta T}$ above gives the classical torsion-function value $0.07367$, i.e.\ an effective conductance $G=\Pabs/\avg{\Delta T}_{\max}\approx4\pi\kappa h$, consistent with the point-source MTF conductance of a drumhead as in Ref.~[K.~Kanellopulos, F.~Ladinig, S.~Emminger, P.~Martini, R.~G.~West, and S.~Schmid, Microsyst.\ Nanoeng.\ \textbf{11}, 28 (2025)]

\subsection{The (2,2) map: numerical example}
\label{sec:SI-map}

As a concrete example, we evaluate $F_{22}$ from Eq.~\eqref{eq:SI-Fseries} (odd indices up to 399, converged to $<10^{-8}$) at three representative positions; the values and their flat-field ratios are listed in Table~\ref{tab:SI-F22}.

\begin{table}[htb]
    \centering
    \caption{Fiber-model responsivity $F_{22}$ and its flat-field ratio at three representative heating positions on an ideal square membrane.}
    \label{tab:SI-F22}
    \begin{tabular}{lccc}
        \hline
         & center $(0.5,0.5)$ & antinode $(0.25,0.25)$ & saddle $(0.5,0.25)$\\
        \hline
        $F_{22}$ (full) & $0.0704$ & $0.0543$ & $0.0605$ \\
        flat-field ratio $F_{22}/A_{\mathrm{env}}-1$ & $-4.5\%$ & $+19.9\%$ & --- \\
        \hline
    \end{tabular}
\end{table}

Three observations:
\begin{enumerate}
    \item The raw map is dominated by the conduction dome: on an ideal square membrane, the global maximum of $F_{22}$ lies at the \emph{center} --- a nodal crossing of the $(2,2)$ mode --- because the envelope Eq.~\eqref{eq:SI-A} peaks there. The mode structure is only a $\pm20\%$ modulation on top of this dome.
    \item Flat-fielding (dividing the raw map by a reference map with no mode structure, e.g.\ a rigid-pad or torsional mode) reveals the mode term directly and simultaneously cancels the local absorptance $a(\mathbf r_0)$.
    \item For trampoline resonators, where the thermal envelope is dominated by a position-independent tether conductance, the dome flattens and the mode term is directly visible in the raw scan --- the regime relevant to the trampoline measurements in this work.
\end{enumerate}

\subsection{Assumptions and limits}
\label{sec:SI-limits}

\textbf{Sanity checks:} (i) every term of Eq.~\eqref{eq:SI-Fseries} vanishes at $x_0,y_0\in\{0,L\}$ (heating the frame does nothing); (ii) in the 1D (string) limit, the responsivity is $\propto\overline{\Delta T}(x_0)$ with no mode-shape contrast; (iii) averaging Eq.~\eqref{eq:SI-Fseries} over $(x_0,y_0)$ reproduces the strain-weighted uniform-heating responsivity, as required by linearity.

\textbf{Assumptions and their validity range:}
\begin{enumerate}
    \item \emph{Linearity:} requires $E\alpha\Delta T\ll\sigma_0$; excludes the buckling-transition regime.
    \item \emph{Quasi-static scanning:} requires dwell time per pixel $\gg$ thermal time constant; otherwise each Fourier component in Eq.~\eqref{eq:SI-T} acquires a low-pass factor $\gamma_{pq}/(\gamma_{pq}+i\omega)$.
    \item \emph{Point-like spot:} a Gaussian spot of $1/e^2$ radius $w_0$ multiplies the $(p,q)$ term by $\approx\exp[-\pi^2(p^2+q^2)w_0^2/(8L^2)]$, smoothing the mode term Eq.~\eqref{eq:SI-Bclosed} on scale $w_0$ --- negligible for $w_0\ll L/2\pi m$.
    \item \emph{No radiation:} valid for $\kappa h/L^2\gg8\varepsilon\sigma_{SB}T_0^3$; for 50\,nm SiN this begins to fail at millimeter side lengths.
    \item \emph{Fiber model:} exact for the oscillatory (mode) response in the interior (Appendix~\ref{app:projector}), but Poisson coupling is absent from the \emph{uniform} component, giving the envelope amplitude a systematic $\mathcal O(30\%)$ band for SiN between the $\nu$-free fiber result and the fully constrained biaxial limit. When fitting measured maps, we therefore float the envelope amplitude and fix the mode-term \emph{shape}, which is $\nu$-free in the exact treatment as well.
\end{enumerate}

\appendix
\section{Exactness of the fiber picture in the interior}
\label{app:projector}

For a Fourier component $T(\mathbf r)=T_{\mathbf q}e^{i\mathbf q\cdot\mathbf r}$ of the temperature field, plane-stress thermoelasticity with equilibrium $\hat q_j\sigma_{ij}=0$ gives
\begin{equation}
  \delta\sigma_{ij}(\mathbf q)
  = -E\alpha\,T_{\mathbf q}\,\big(\delta_{ij} - \hat q_i\hat q_j\big).
  \label{eq:SI-projector}
\end{equation}
Each temperature component relieves stress \emph{only transverse to its own wavevector} --- precisely the fiber statement: a temperature variation along $x$ cannot modulate $\sigma_{xx}$, but fully relieves $\sigma_{yy}$ of the crossing fibers in proportion to the mean over their length. Projecting Eq.~\eqref{eq:SI-projector} onto the coordinate axes and averaging reproduces Eq.~\eqref{eq:SI-fiber} exactly. Furthermore, the diagonal Fourier components of the mode strain tensor at $(\pm2a,\pm2b)$ --- which would generate a term $\propto-\phi^2(x_0,y_0)$ in the naive local model --- contract with the projector in Eq.~\eqref{eq:SI-projector} to exactly zero; only the stripe components at $(\pm2a,0)$ and $(0,\pm2b)$ survive, reproducing the separable structure of Eq.~\eqref{eq:SI-Bclosed}. The fiber model and the exact plane-stress solution agree term by term in the interior; they can differ at $\mathcal O(1)$ only in the boundary layers (the windows $w_c$) and in the uniform component (Sec.~\ref{sec:SI-limits}).

\section{Proofs of Lemmas 1 and 2}
\label{app:lemmas}

\textit{Lemma 1:} since $(\tfrac{\dd^2}{\dd\theta^2}+4m^2)\sin q\theta = (4m^2-q^2)\sin q\theta$, the sum $g(\theta)=\sum_{q\,\mathrm{odd}}\sin(q\theta)/[q(4m^2-q^2)]$ satisfies $g''+4m^2g=\sum_{q\,\mathrm{odd}}\sin(q\theta)/q=\pi/4$ (square-wave series). The general solution consistent with $g(0)=0$ and symmetry about $\theta=\pi/2$ is $g(\theta)=\frac{\pi}{8m^2}\sin^2(m\theta)$.

\textit{Lemma 2:} analogously, with $(c^2-\tfrac{\dd^2}{\dd\theta^2})$, the sum $\sum_{p\,\mathrm{odd}}\sin(p\theta)/[p(p^2+c^2)]$ satisfies $-g''+c^2g=\pi/4$; the symmetric solution vanishing at $\theta=0,\pi$ is $g(\theta)=\frac{\pi}{4c^2}\big[1-\cosh(c(\theta-\pi/2))/\cosh(c\pi/2)\big]$.

\section{Elementary integrals used in the assembly step}
\label{app:integrals}

Using $\frac{1}{L}\int_0^L\sin(p\pi x/L)\,\dd x = \frac{2}{p\pi}$ for odd $p$ (zero for even $p$), the fiber-mean temperature is
\begin{equation}
  \overline{\Delta T}_x(y)
  = \frac{8\Pabs}{\pi^3\kappa h}
    \sum_{\substack{p\;\mathrm{odd}\\ q\ge1}}
    \frac{\sin\frac{q\pi y}{L}\;\sin\frac{p\pi x_0}{L}\sin\frac{q\pi y_0}{L}}
         {p\,(p^2+q^2)}.
  \label{eq:SI-Tbar}
\end{equation}
Using $\sin^2(m\pi y/L)=\tfrac12[1-\cos(2m\pi y/L)]$ and product-to-sum identities, the overlap integral for odd $q$ (even $q$ vanishes) is
\begin{equation}
  I_q \equiv \int_0^L \sin\frac{q\pi y}{L}\,\sin^2\frac{m\pi y}{L}\,\dd y
  = \frac{L}{\pi}\,\frac{4m^2}{q\,(4m^2-q^2)},
  \label{eq:SI-Iq}
\end{equation}
which has no resonant term since $q$ is odd and $2m$ even, and is largest for small odd $q$ (the slowly varying temperature components couple best to the modal loading).

\section{Numerical evaluation}
\label{app:numerics}

The responsivity maps were produced by direct summation of Eq.~\eqref{eq:SI-Fseries} with odd $p,q\le399$ (vectorized as a matrix product $S_p^{\mathsf T}CS_q$; $\sim0.1$\,s on a laptop). The decomposition $F=A+B$ using Eqs.~\eqref{eq:SI-Bclosed} and \eqref{eq:SI-A} was verified to $4\times10^{-9}$ (truncation level); the closed form Eq.~\eqref{eq:SI-Bclosed} matches the series to five digits. A self-contained Python script implementing this evaluation accompanies the Supplementary Material.

\section{Linearized radiative extension of the model}
\label{sec:SI-radiation}

The main-text model neglects radiative heat loss. Here we show how radiation enters and why it produces only a small, smooth flattening of the response for the membranes studied here.

Adding radiative loss from both faces, linearized about the frame temperature $T_0$ (valid for $\Delta T \ll T_0$), the steady-state heat equation Eq.~\eqref{eq:SI-heat} becomes screened,
\begin{equation}
  -\kappa h\,\nabla^2 \Delta T \;+\; 8\,\varepsilon\sigma_{\mathrm{SB}}T_0^3\,\Delta T
  \;=\; \Pabs\,\delta(x-x_0)\,\delta(y-y_0),
  \label{eq:SI-heat-rad}
\end{equation}
with $\varepsilon$ the emissivity and $\sigma_{\mathrm{SB}}$ the Stefan--Boltzmann constant. In the Dirichlet basis this shifts each denominator of the temperature Green's function Eq.~\eqref{eq:SI-T} by a constant,
\begin{equation}
  \frac{1}{p^2+q^2} \;\longrightarrow\; \frac{1}{p^2+q^2+\gamma},
  \qquad
  \gamma \equiv \frac{8\,\varepsilon\sigma_{\mathrm{SB}}T_0^3\,L^2}{\pi^2\,\kappa h}.
  \label{eq:SI-gamma}
\end{equation}
Because every downstream step (fiber averaging, strain-overlap integrals) is linear in the temperature field, the same substitution carries through unchanged into $F_{nm}$, $A_{nm}$ and $B_{nm}$: radiation is implemented by the single replacement Eq.~\eqref{eq:SI-gamma}.

The effect is hierarchical. The screening suppresses each Fourier component by $1/[1+\gamma/(p^2+q^2)]$, so it acts most strongly on the lowest, broadest component --- the one that dominates the conduction dome --- and negligibly on the sharp, high-order components that carry the mode-specific structure. For the present devices ($\varepsilon = 0.05$, $T_0 = 295\,$K, $L = 500\,$\textmu m, $h = 50\,$nm, $\kappa \approx 3\,$W\,m$^{-1}$K$^{-1}$) one finds $\gamma \approx 0.10$, i.e.\ a $\sim\!5\%$ suppression of the $(1,1)$ temperature component and $\lesssim\!1\%$ for higher components. Radiation therefore flattens the envelope $A$ at the few-percent level while leaving the mode term $B$ essentially unchanged; the exact decomposition $F=A+B$ survives, with Lemma~1 unaffected and the window parameter in Lemma~2 shifted from $2m$ to $\sqrt{4m^2+\gamma}$ --- a sub-percent change. This magnitude and shape match the $\sim\!1\%$ dome flattening seen in the data (main text), which is why we retain the conduction-only model in the main text and treat radiation only as this small, understood correction.

\section{Data processing: tile stitching and mode-term extraction}
\label{sec:SI-pipeline}

The full-membrane maps of Fig.~\ref{fig:membranes} are assembled from five overlapping $300\times300$\,\textmu m scan tiles per mode (a central tile, 00, and four corner tiles, 01--04), each a regular $21\times21$ grid at $15$\,\textmu m pitch. The physical stage offset between tiles is not recorded in the data files, so the tiles are registered against the fiber model itself.

\emph{Tile averaging and artifact rejection.} Where a tile was scanned more than once, the repeats are averaged before registration. One scan was excluded: an early $515$\,kHz corner-04 scan showed a lock-in/mode-hopping artifact (a spurious maximum of ${\sim}8\times10^{5}$\,Hz against a physical signal ceiling of ${\sim}1800$\,Hz, and a correlation of only $0.42$ with the valid repeat of the same tile --- inconsistent with measurement noise), and was discarded in favour of the clean scan.

\emph{Center placement.} The central tile 00 defines the coordinate frame and is never geometrically adjusted; its own scan window, however, need not be centerd on the membrane. Since the $325$\,kHz $(1,1)$ mode has a single antinode at the geometric center, the membrane center is located by fitting the theoretical $(1,1)$ responsivity --- the diffusion-weighted shape of Eq.~\eqref{eq:SI-Fseries}, not a naive $\phi^2$ profile --- to all $\sim\!441$ points of the tile-00 scan, with the center offset, an overall amplitude and a constant offset as free parameters. This uses the full scan rather than a single-pixel maximum, which would be limited by the $15$\,\textmu m grid pitch.

\emph{Corner-tile registration.} Each corner tile is placed by a translation $(\delta x,\delta y)$ that maximizes agreement with the already-placed tiles in the overlap region; the scan-window size is fixed, so no scaling of coordinates is applied. A small multiplicative correction (constrained to $\pm5\%$) is solved in closed form to absorb frequency-calibration drift between scan sessions. Because all four modes were scanned at the \emph{same} physical stage positions, the translation must be common to all modes: it is fitted independently per mode, the four values are checked for consistency (their spread is a diagnostic for a questionable placement), and each tile is then frozen at the mode-averaged position with only the per-mode amplitude refit. Registration quality is quantified by the Pearson correlation and RMS residual in each overlap region.

\emph{Merging.} All five tiles' points are concatenated into a single point cloud and interpolated onto one regular grid per mode (Delaunay triangulation), so overlapping tiles are blended rather than one selected over another. The map is cropped to the nominal $500\times500$\,\textmu m membrane. We use the nominal $L = 500\,$\textmu m throughout the model; the fitted membrane size agrees with this to within ${\sim}2\%$, and the mode-term shape is insensitive to a change of this order.

\emph{Mode-term extraction.} Each stitched raw map is normalized by the resonance frequency to give $|\delta f|/f_0$. A single overall amplitude $k$ is fitted against the theoretical mode shape $F_{nm}$ (least squares over all valid grid points), after which the mode term is displayed as $B_{nm}/A_\mathrm{max} = (|\delta f|/f_0)/(k\,A_\mathrm{max}) - A_\mathrm{env}/A_\mathrm{max}$, i.e.\ the envelope-subtracted response normalized by the single peak-envelope constant $A_\mathrm{max}$ (Sec.~\ref{sec:theory}). Only this one amplitude is floated; the spatial shape is parameter-free. Normalizing by the constant $A_\mathrm{max}$ rather than pointwise by $A_\mathrm{env}$ avoids the noise amplification of a ratio near the clamped edge, where $A_\mathrm{env}\!\to\!0$, and preserves the true amplitude of the mode contribution across the map. The absorbed-power calibration and any per-pixel absorptance cancel in this construction up to the overall constant $k$.

\end{document}